\documentclass[12pt,preprint]{aastex}

\newcommand{\myemail}{zhekovs@colorado.edu}
\newcommand{\kms}{~km s$^{-1}$~}
\newcommand{\WR}{WR~147~}
\newcommand{\WRN}{WR~147N~}
\newcommand{\WRS}{WR~147S~}
\newcommand{\dotM}{~M$_{\odot}$~yr$^{-1}$~}

\shorttitle{X-rays from \WR}
\shortauthors{Zhekov \& Park}

\begin{document}


\title{
{\it Chandra} HETG Observations of the 
Colliding Stellar Wind System \WR 
}



\author{Svetozar A. Zhekov\altaffilmark{1,3}
and Sangwook Park\altaffilmark{2} }

\altaffiltext{1}{JILA, University of Colorado, Boulder, CO
80309-0440, USA; \myemail}
\altaffiltext{2}{Department of Astronomy and Astrophysics,
Pennsylvania State University, 525 Davey Laboratory, University
Park, PA 16802, USA; park@astro.psu.edu}
\altaffiltext{3}{On leave from Space Research Institute, Sofia,
Bulgaria}


\begin{abstract}
We present an extended analysis of  deep {\it Chandra} HETG 
observations of the WR$+$OB binary system \WR that was resolved into a
double X-ray source \citep{zhp_10}.
Our analysis of the profiles of strong emission lines shows that their
centroids are blue-shifted in the spectrum of the northern X-ray
source. We find no suppressed forbidden line in the He-like triplets
which indicates that the X-ray emitting region is not located near
enough to the stars in the binary system to be significantly affected
by their UV radiation.
The most likely physical picture that emerges from the entire set of
HETG data suggests that the northern X-ray source can be associated 
with the colliding stellar wind region in the wide WR$+$OB binary 
system,
while the X-rays of its southern counterpart, the WN8 star, are
result from  stellar wind shocking onto a close companion
(a hypothesized third star in the system).

\end{abstract}





\section{Introduction}
WR$+$OB binaries are the brightest X-ray sources amongst the
Wolf-Rayet (WR) stars \citep{po_87}. Their enhanced emission 
originates from the interaction region of the winds of the two 
massive stars (\citealt{pril_76}; \citealt{cherep_76}). 
Since the winds are highly supersonic, the interaction 
region is bounded by two shocks each compressing the stellar wind of 
the WR or OB star, and a contact discontinuity surface separating 
their shocked plasmas (for the first hydrodynamic models see 
\citealt{lm_90};
\citealt{luo_90}; \citealt{st_92}; Myasnikov \& Zhekov 1991, 1993). 
Given the wind velocities (typical values of 1,000-3,000\kms), 
the postshock temperatures are in the keV range, and  most of 
the plasma emission is thus in X-rays. 
If the shocks are adiabatic, the X-ray luminosity of the
colliding stellar winds (CSW) is proportional to the square of the
mass-loss rate ($\dot{M}$) and inversely to the binary separation
($D$):
$L_X \propto \dot{M}^2 V_{wind}^{-3.2} D^{-1}$ (see the references
above for discussion). From this follows:
(i) the shocked WR wind dominates the X-ray emission from CSW in
WR$+$OB binaries due to its much more massive wind ($\dot{M}$ of
a WR star is about an order of magnitude higher than that of an O
star);
(ii)  by the same argument, a WR$+$OB binary will be more
luminous in X-rays than an OB$+$OB binary, with similar
wind velocities and binary separation.

On the other hand, CSWs in close and wide binaries have quite
different behavior. While in the former, shocks are radiative
and thus the interaction region is subject to instabilities (e.g.,
\citealt{st_92}; \citealt{mzhb_98}; \citealt{wa_00}), the shocks are
adiabatic in the latter and the effects of thermal conduction
might also be important \citep{mzh_98}. Moreover, the
effects of electron-ion temperature equilibration behind the
shocks \citep{zhsk_00} and non-equilibrium ionization
\citep{zh_07} can influence the X-ray emission from CSWs.

A difference between CSWs in close and wide WR binaries is also seen
in radio, and as a rule non-thermal radio (NTR) sources
are associated with the wide binary systems \citep{do_00}. 
The strong shocks are the likely place
for accelerating relativistic particles, and the CSWs in wide
binaries offer good conditions for this mechanism to operate
efficiently since the shocks are located relatively far from the
optically bright sources (the stars) and thus the inverse
Compton losses
are minimal (for the first detailed models of NTR emission from
CSWs see \citealt{do_03}; \citealt{pitt_06}).

It is therefore seen that the CSWs in stellar binaries are an
ideal laboratory for studying the wealth of physical processes
related to strong shocks. With the launch of the modern X-ray
observatories providing high spectral and spatial resolution
observations ({\it Chandra}, {\it XMM-Newton}),
a new window has opened for gathering detailed information about
this exciting phenomenon. By confronting theoretical models of
CSWs with observations, we can rigorously test them, and thus
improve our understanding of the underlying physics. For such a goal, 
it is necessary to have an object which is bright enough in X-rays,
it is a strong non-thermal radio source and
it is relatively close to us. Thus, its radio-to-X-ray emission might
be spatially resolved and provide us with detailed information about
the CSW region and the stars in the binary system.
Because the two components in the binary system can be separated, 
\WR is unique amongst the massive WR$+$OB
binaries and offers a rare opportunity for studying the CSW-binary
phenomenon in its entirety.

We report here results from our analysis of deep {\it Chandra} HETG
observations of the CSW binary \WR. 
\citet{zhp_10} presented the first part of our study based on the
zeroth-order HETG data whose most important result is that the X-ray
emission from \WR was resolved into two sources. 
This paper is organized as follows.
We give basic 
information about the WR$+$OB
binary \WR in \ Section \ref{sec:thesystem}. In  Section 
\ref{sec:observations}, we briefly review the {\it Chandra} HETG 
observations.
In Section \ref{sec:lines}, we present the results from analysis of
strong X-ray emission lines. In Section \ref{sec:origin}, we discuss
the origin of X-rays in the two X-ray components of \WR.
In Section \ref{sec:global}, we report results from the global
spectral models. In Section \ref{sec:discussion}, we discuss our
results and we list our conclusions in Section \ref{sec:conclusions}.

\section{The Wolf-Rayet Binary \WR}
\label{sec:thesystem}
The Wolf-Rayet star \WR (WR$+$OB; \citealt{vdh_01}) is a classical
example of colliding wind binary at a distance of $630\pm70$ pc 
\citep{ch_92}.
High-resolution radio observations resolved its emission into two 
components: a southern thermal source, \WRS (the WN8 star in the 
system), and a northern non-thermal source, \WRN 
(\citealt{ab_86}; \citealt{mo_89}; \citealt{ch_92};
\citealt{con_96}; \citealt{wi_97}; \citealt{sk_99})
with separation of $\sim 0\farcs57$.
The binary system was spatially resolved both in infrared and optical
to have a separation of $\sim 0\farcs64$ (\citealt{wi_97};
\citealt{nie_98})
which at the distance to this
object corresponds to projected (or minimum) binary separation of
$403\pm13$ au.
While the spectral type of the WR star in the binary is well defined
(WN8h; \citealt{vdh_01}), that of the OB companion is not well
constrained. 
From spatially resolved  near-infrared and optical photometry,
the spectral type of the latter was estimated correspondingly 
as a B0.5V \citep{wi_97} and O8-9 V-III  \citep{nie_98}. And
\citet{lepine_01} classified it as a O5-7 I-II from spatially 
resolved spectra, but in the red optical domain.

Due to the high extinction towards \WR (A$_V=10.45$; A$_V=$
A$_v/1.11$; A$_v=11.6$; \citealt{vdh_01}), the WR wind parameters,
$\dot{M}=4\times10^{-5}$\dotM; $V_{wind}=950$\kms, were derived from 
radio and NIR
observations (\citealt{sk_99}; \citealt{mo_00}). For consistency with
the previous works (\citealt{sk_99}; \citealt{sk_07};
\citealt{zh_07}), we adopt $\dot{M}=6.6\times10^{-7}$\dotM;
$V_{wind}=1600$\kms for the stellar wind parameters of the OB
companion.

The previous X-ray observations of \WR have revealed the presence of
thermal emission from high temperature plasma: $\mbox{kT} \geq
0.5$~keV ({\it Einstein} observatory; \citealt{cai_85});
$\mbox{kT} \approx 1$ keV ({\it ASCA}; \citealt{sk_99});
$\mbox{kT} = 2.7$ keV ({\it XMM-Newton}; \citealt{sk_07})
as the {\it XMM-Newton} observations detected the Fe K$_{\alpha}$
complex at 6.67 keV which is a clear sign of thermal X-rays even at
high energies.
Note that the temperature change simply reflects the improving quality
of the X-ray data over the years.

The observations with {\it Chandra} High-Resolution Camera (having low 
photon statistics, $\sim 148$ source counts, and no spectral 
information) found indications that the X-ray emitting region in \WR 
might be extended and it peaks north from the WN8 star although a 
deeper X-ray image was needed to determine the degree of spatial 
extent \citep{pitt_02}.

The recent high resolution {\it Chandra} observations resolved
the X-ray emission from \WR into two sources, \WRN and
\WRS, with a spatial separation of $\approx 0\farcs60$ \citep{zhp_10}.
The corresponding analysis of undispersed spectra showed that \WRN and
\WRS have different global characteristics as the latter being more
absorbed and having higher plasma temperature: 
N$_H = 2.28\,[2.08 - 2.57]\times10^{22}$ cm$^{-2}$; 
$\mbox{kT} = 1.78\,[1.52 - 1.98]$ keV 
for \WRN;
N$_H = 3.83\,[3.51 - 4.20]\times10^{22}$ cm$^{-2}$; 
$\mbox{kT} = 2.36\,[2.12 - 2.56]$ keV 
for \WRS.
It is worth noting that the absorption towards \WRN almost perfectly
corresponds to the optical extinction of \WR if the \citet{go_75}
conversion is used.

\section{Observations and Data Reduction}
\label{sec:observations}
\WR was observed with {\it Chandra} in the configuration  HETG-ACIS-S
in eight consecutive runs
(\dataset[ADS/Sa.CXO\#obs/09941]{Chandra ObsIds: 9941,}
\dataset[ADS/Sa.CXO\#obs/09942]{ 9942,}
\dataset[ADS/Sa.CXO\#obs/10675]{10675,}
\dataset[ADS/Sa.CXO\#obs/10676]{10676,}
\dataset[ADS/Sa.CXO\#obs/10677]{10677,}
\dataset[ADS/Sa.CXO\#obs/10678]{10678,}
\dataset[ADS/Sa.CXO\#obs/10893]{10893 and}
\dataset[ADS/Sa.CXO\#obs/10897]{10897})
in the period 2009 Mar 28 - Apr 10, providing
a total effective exposure of 286 ksec.
The roll angle was between $75^{\circ}$ and $82^{\circ}$: therefore,
the dispersion axis was aligned approximately with the position angle
of the binary system \WR as derived in the optical
(P.A. $=350^{\circ}\pm2$; \citealt{nie_98}).
The instrument configuration was such that the {\it negative}
first-order MEG/HEG arms were pointing to north.
By default, the pixel randomization is switched off in grating data.

As reported in \citet{zhp_10}, the X-ray emission from \WR was 
spatially resolved 
into a northern, \WRN, and a southern counterpart, \WRS (the WN8 star 
in the binary). Thus, 
following the Science Threads for Grating
Spectroscopy in the CIAO 4.1.2 \footnote{Chandra Interactive Analysis
of Observations
(CIAO), http://cxc.harvard.edu/ciao/} data analysis software,
the positive
and negative first-order MEG/HEG spectra for each of the eight
observations were extracted centered on the \WRS position on the sky.
The {\it Chandra} calibration database CALDB v.4.1.3 was used in this
analysis.
The resultant spectra were merged into one spectrum each for the
positive and negative MEG/HEG arms with respective total counts of
2742 (MEG$+1$), 2396 (MEG$-1$),
1438 (HEG$+1$) and 1606 (HEG$-1$).
The zeroth-order data were discussed in \citet{zhp_10} and we only
mention that the total number of zero order counts in the \WRN and 
\WRS spectra were 2158 and 5108, respectively.
We note that terms \WR and WR~147N$+$S will be used throughout the
text to refer to the total X-ray emission (and spectra) of the studied
WR$+$OB binary system.

For the spectral analysis in this study, we made use of standard as
well as custom models in version 11.3.2 of XSPEC
\citep{Arnaud96}.





\section{Spectral Lines}
\label{sec:lines}
The X-ray emission from \WR is dominated by \WRS \citep{zhp_10} 
and the presence of another
source (\WRN) at $\sim 0\farcs6-0\farcs64$ will cause the
emission
line profiles to be `blurred' since the \WRS - \WRN orientation on
the sky is approximately along the South-North direction,
that is parallel to the dispersion axis.
If there were no line shifts in the \WRN emission, the line centroids
in the total \WR spectrum
would correspondingly show {\it blue shifts} in the positive and
{\it red shifts} in the negative arms of the MEG/HEG spectra and
their values will be the same
($\vert z(+1)\vert = \vert z(-1)\vert$).
If the \WRN emission was intrinsically blue-shifted, this would result
in larger line shifts in the MEG/HEG($+1$) than in the
MEG/HEG($-1$) for each spectral line
($\vert z(+1)\vert > \vert z(-1)\vert$).
The result would be just the opposite
($\vert z(+1)\vert < \vert z(-1)\vert$),
if the \WRN spectrum was red-shifted.

\begin{figure}[hp]
\centering\includegraphics[width=2.80in,height=2.0in]{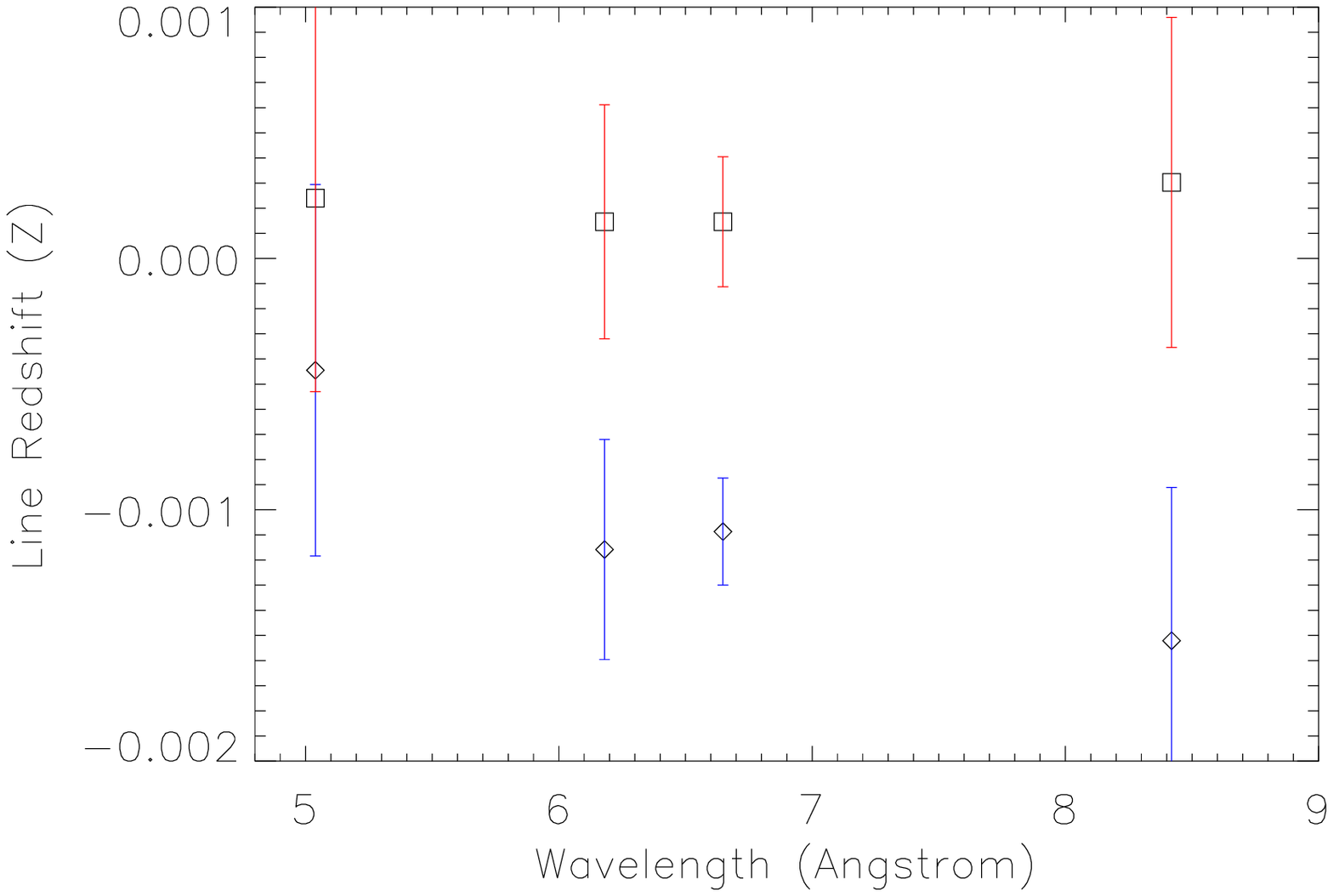}
\caption{
Line shifts 
of S XV (5.04\AA), Si XIV (6.18\AA), Si XIII (6.65\AA) and Mg XII
(8.42\AA) in the total MEG/HEG($+1$) and ($-1$) spectra.
The results for the positive, $z(+1)$, and negative, $z(-1)$, arm are
correspondingly denoted by diamonds (with error bars in blue) and
squares (with error bars in red).
For each line, the MEG and HEG data were fitted simultaneously.
}
\label{fig:shifts}
\end{figure}

For each spectral line, we fitted all four spectra, MEG($+1/-1$) and
HEG($+1/-1$), simultaneously as they were re-binned every two bins to
improve the photon statistics.
For the S XV and Si XIII He-like triplets, we fitted a sum of three
Gaussians and a constant continuum. The centers of the
triplet components were held fixed according to the {\it Chandra} atomic
data base\footnote{For ATOMDB, see http://cxc.harvard.edu/atomdb/} 
and all components
share the same line width and line shifts. 
Similarly, we fitted the Si XIV and Mg XII H-like doublets but the
component intensities were fixed through their atomic data values.
Figure~\ref{fig:shifts} shows the results for the line shifts of
prominent lines in the X-ray spectrum of \WR.
We see that all the lines are blue-shifted in the MEG/HEG($+1$) and
red-shifted in the MEG/HEG($-1$) spectra and $\vert z(+1)\vert > \vert
z(-1)\vert$. Thus, the results indicate {\it blue-shifted}
X-ray emission from \WRN.
And, we note that we do not find suppressed forbidden line in the
He-like triplets.

\begin{figure}[hp]
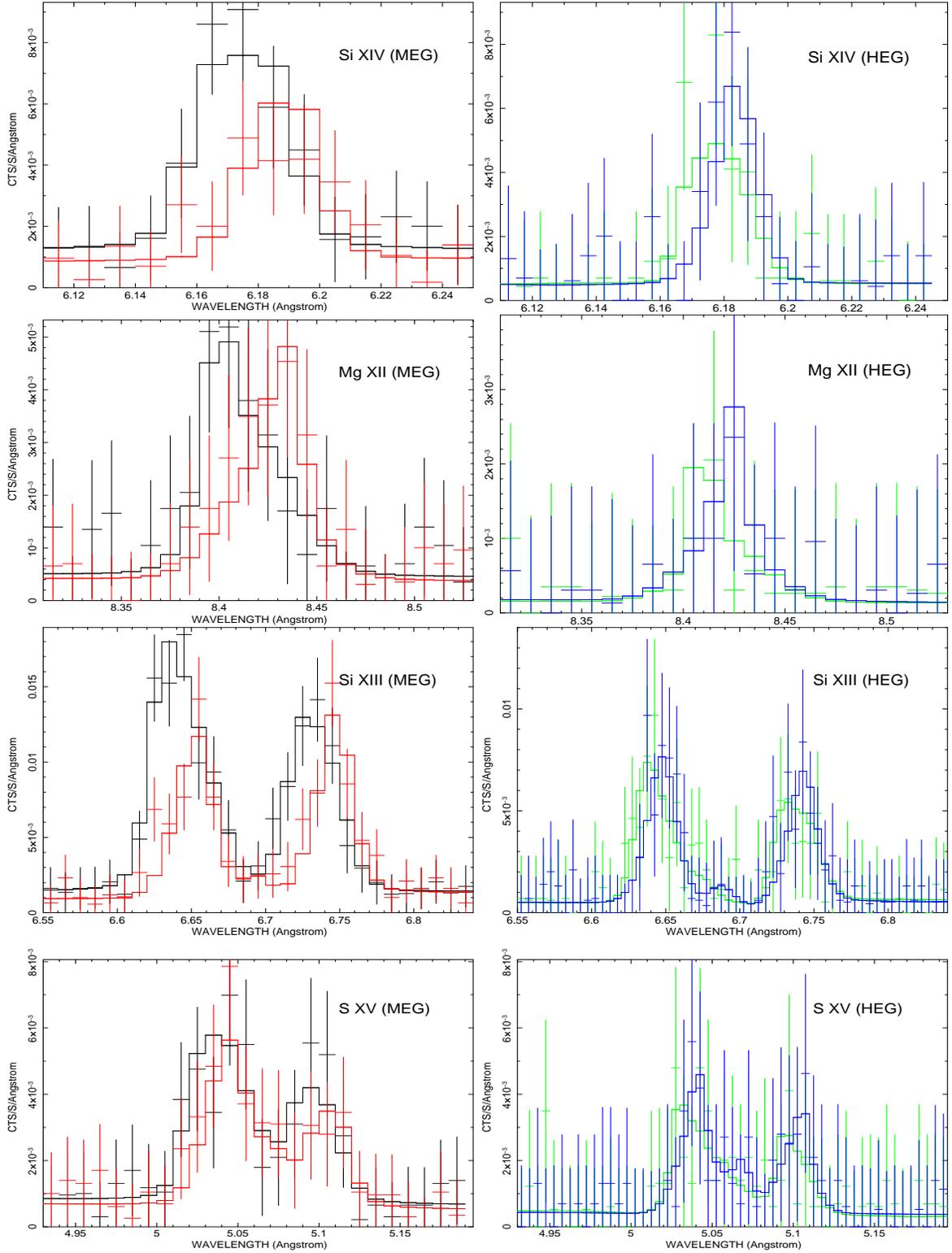

 \centering\includegraphics[width=2.in,height=3.in,angle=-90]{f2a.eps}
 \centering\includegraphics[width=2.in,height=3.in,angle=-90]{f2b.eps}
 \centering\includegraphics[width=2.in,height=3.in,angle=-90]{f2c.eps}
 \centering\includegraphics[width=2.in,height=3.in,angle=-90]{f2d.eps}
 \centering\includegraphics[width=2.in,height=3.in,angle=-90]{f2e.eps}
 \centering\includegraphics[width=2.in,height=3.in,angle=-90]{f2f.eps}
 \centering\includegraphics[width=2.in,height=3.in,angle=-90]{f2g.eps}
 \centering\includegraphics[width=2.in,height=3.in,angle=-90]{f2h.eps}
\caption{
Line profile fits with the two-component model.
{\it Left panel} shows the MEG($+1$) in black and the MEG($-1$)
in red.
{\it Right panel}  shows the HEG($+1$) in green and the HEG($-1$)
in blue.
}
\label{fig:profiles}
\end{figure}

Motivated by these results, we developed a custom model for XSPEC to
fit the line profiles that consist of line emission from two sources
with an `offset', $\Delta$, for the line center of the
second one. 
We note that for the instrument configuration of our observations
(\S~\ref{sec:observations}) and since the second source (\WRN) is 
located north from \WRS
its spectral lines will get a red shift in the negative first-order
spectra and a blue shift in the positive first-order spectra that
correspond to the spatial offset $\Delta$. But for MEG and HEG this 
line shift will be different in units of wavelength due to their 
different spectral resolution ($\Delta_{MEG} = 2 \Delta_{HEG}$).
For each source, the profiles of the spectral line doublets 
and triplets were treated as described above.
We fitted this model to the line profiles of the H-like
doublets of Mg XII and Si XIV as well as of the He-like triplets of
Si XIII and S XV again simultaneously for the four first-order
spectra. 
The fit results are given in Table~\ref{tab:lines} and 
individual line profiles are shown in Fig.~\ref{fig:profiles}.

From the fit results we see that the line profiles in the spectrum of 
\WRS are broader than those in \WRN: bulk gas velocities (FWHM) with a
typical value of $\approx 1000$\kms are present in the former while
the line widths indicate slower gas motion ($ < 1000$\kms)  for the 
latter. This may be a sign of different X-ray production mechanisms
and we will return to this issue in \S~\ref{sec:origin}.
An important feature of these mechanisms is that we do not find
suppressed forbidden line in the He-like triplets which means that
the X-ray emission plasma has relatively low density and/or the hot
plasma regions both in \WRS and \WRN are located far
enough from strong UV sources.

Perhaps  the two most interesting fit results are related to 
\WRN alone. First, we see that despite their relatively large errors
the centroids of {\it all} the studied lines are {\it blue-shifted} 
as anticipated from the preliminary analysis of the line profiles 
(see above). Second, the values for the spatial `offset' between \WRN 
and  \WRS are consistent between different spectral lines. 
Interestingly, the average value from all four spectral lines, 
$\bar{\Delta} = 0\farcs603^{+0.10}_{-0.08}$, is in a
very good correspondence  with the source separation measured directly 
from the deconvolved (1.0 - 2.0 keV) zeroth-order image of \WR
\citep{zhp_10}. 
This is a very important internal cross-check for the HETG results and
another nice illustration of the superior spatial and spectral
resolution of the {\it Chandra} observatory.

Finally, for the analysis here we assumed that the spectral lines of
\WRS are not shifted and the reason was that the fits are too
complicated already, thus, an extra free parameter in the fits cannot
be constrained well due to the quality of the data. 
But we note that we derived results generally consistent with zero line
shifts if we let this parameter vary
(the average value from the line shifts of the prominent line complexes 
in the spectrum of \WRS is -96$^{+75}_{-76}$\kms).

\section{The Origin of X-ray Emission from \WR}
\label{sec:origin}

Thanks to its superior spatial resolution, {\it Chandra} resolved the
CSW binary \WR into a double X-ray source. The different spectral
characteristics (plasma temperature and X-ray absorption) as deduced
from the analysis of the undispersed spectra of both components 
(\citealt{zhp_10}; see also \S~\ref{sec:thesystem}) as well as the 
results from the fits to  spectral line profiles in the first-order 
spectra (\S~\ref{sec:lines}) likely point to different emission 
mechanisms responsible for the X-ray emission from \WRN and \WRS.

\subsection{X-rays from \WRN}
\label{subsec:wr147n}
Since \WR is a spatially resolved radio source with thermal and
non-thermal components, all the
analyses of its X-ray emission before the {\it Chandra} observations 
have assumed that the colliding stellar winds are
responsible for the total X-ray emission from this WR$+$OB binary. 
We now know that the physical picture is more complex and at least
two components are contributing to the X-ray emission of the
\WR system. Thus, only the northern source, \WRN, can be
associated with the CSW region in this binary system \citep{zhp_10}.
But, the presence of another massive star (the OB companion in the
binary) in vicinity of \WRN requires a more careful look into such an
identification.

We recall that \citet{zhp_10} found that the CSW model with nominal
wind parameters for \WR perfectly matches the shape of the X-ray
spectrum of \WRN. They also reported
a 
correspondence between the
location of \WRN in X-rays and \WRN in the radio, and the latter is
definitely associated with the CSW region in \WR \citep{con_99}.
We could prove that \WRN is located in the CSW region of the
binary system if the X-ray 
separation  between \WRN and \WRS is smaller
than the value derived from the optical and NIR imaging. This would
indicate that the northern X-ray source is detached from the surface
of the OB companion.
However, the values derived from the X-ray analysis 
do not
provide solid evidence for that due to their uncertainties 
(see \S~\ref{sec:lines} and Table~\ref{tab:lines}).

Extended emission in \WRN would support that this source is the CSW
region, 
but we  see no extended X-ray emission in the HETG zeroth-order image
that might have morphology similar to that of the extended non-thermal
radio emission in \WRN \citep{con_99}.
A reason for that could be the limited photon statistics in soft
X-rays (only $\sim 700$ counts in the 1-2 keV band) and we note that 
soft X-rays originate downstream from the shocks that is in the
`outskirts' of the CSW region. 
Thus,  future X-ray observations that
provide images with higher quality may help us establish with
certainty the morphology of \WRN which in turn will be very helpful
in resolving the issue about the proper identification of this X-ray
source.

On the other hand, the massive OB stars are X-ray sources and
could it be that the northern X-ray source, \WRN, is in fact the OB
companion in this wide binary system? In general, this might well be
the case but there are some pieces of indirect evidence that make
such an identification more unlikely. For example, the plasma
temperature in \WRN is relatively high ($\mbox{kT}= 1.78$ keV;
\S~\ref{sec:thesystem}) while OB stars are in general 'soft' X-ray sources 
with temperatures below 1 keV (e.g., \citealt{woj_05};
\citealt{zhp_07}; see also \S 4.1.2 and \S 4.3 in the review paper
of \citealt{gudel_naze_09} and the references therein).
As a reference case, we ran XSPEC simulations having exposure of 286
ksec for a `soft' X-ray source with $\mbox{kT} = 0.6$~keV, 
L$_X (0.5-10 \mbox{ keV}) = 10^{31}$ ergs s$^{-1}$ 
and N$_H = 2.3\times10^{22}$~cm$^{-2}$ (typical for \WRN), using 
the ancillary response functions from our zeroth-order HETG 
observation. We found 30-40 source counts or $< 2$\% from the
total zero order counts of \WRN. This indicates that a `soft'
X-ray source will likely have a small contribution to the total
emission from \WRN.
But, exceptions could be the rare hot magnetic objects like the massive 
O star $\theta^1$ Ori C and the B star $\tau$ Sco whose spectra show 
signs of considerably hotter plasma (\citealt{sch_03};
\citealt{gagne_05}; \citealt{cohen_03}).
The X-ray production mechanism that is believed to operate in such
objects is that of magnetically confined wind shocks
(MCWS, e.g., \citealt{babel_97}; \citealt{ud_02}).
This model suggests that the X-ray emission mostly originates in
regions very close to the stellar surface, thus, due to the strong
stellar UV emission the forbidden line can likely be suppressed even 
in the He-like triplets as Si XIII and S XV. In fact, this is the case 
in the classical MCWS object $\theta^1$ Ori C \citep{gagne_05} and 
similar indications are found for $\tau$ Sco \citep{cohen_03}.
Opposite to this, we do not find suppressed forbidden line in the Si
XIII and S XV He-like triplets in the X-ray spectrum of \WRN
(\S~\ref{sec:lines}) which indicates that these lines are formed in
regions far from the surface of the OB companion.
It is also worth noting that because of the expected 
decay of the magnetic field strength with the age of a massive star,
the MCWS mechanism is associated only with young massive stars 
(age $\le 1$ Myr; \citealt{sch_03}). But, the age of the OB star in 
the wide WR$+$OB binary system \WR must be larger than 1-2 Myr 
as indicated by the presence of a WR star in the system.

From all this, a physical picture in which \WRN resides in the CSW 
region seems more realistic than  assuming that \WRN is associated 
with the OB companion in the massive binary system \WR. 
However, we should keep in mind that the CSW model with nominal wind
parameters for \WR overestimates the observed \WRN luminosity by a 
factor of $\sim 16$ \citep{zhp_10}, a discrepancy that must find
reasonable explanation in the CSW scenario. We will discuss 
possible solutions to this problem in \S~\ref{sec:discussion}.

\subsection{\WRS: Unusual X-ray Source}
\label{subsec:wr147s}
The X-ray detection of the WN8 star in \WR and its brightness make it 
unusual amongst the WR stars of the same subtype. As shown by
\citet{sk_10}, single WN7-9 stars are very
weak X-ray sources. And, the X-ray emission of \WRS is quite hard
as the analysis of its undispersed spectrum
indicated ($\mbox{kT} = 2.36$~keV).

Skinner et al. (2002a,b, 2010) discussed in some detail possible
mechanisms for X-ray production in WN stars, which include:
instability-driven wind shocks;
magnetically confined wind shocks; wind
accretion shocks; colliding wind shocks (including the case of
stellar wind shocking onto a close companion);
non-thermal X-ray emission.
We note that none of these mechanisms finds solid observational
support for the moment and
each of them has its own limitations and caveats.
We recall that instability-driven wind shocks are supposed to be soft
X-ray emitters which is not the case with \WRS. Accretion wind shocks
result in a relatively high X-ray luminosity ($\sim 10^{36-37}$~ergs
s$^{-1}$) and the observed value is short by 3-4 orders of magnitude.
As discussed above (\S~\ref{subsec:wr147n}), in the magnetically 
confined wind shock picture X-rays  form close to the stellar 
surface which results in a suppressed forbidden line even in He-like 
triplets as Si XIII and S XV due to the strong stellar UV emission. 
Opposite to this, 
we find no indications for such a suppression in the spectrum of \WRS
(\S~\ref{sec:lines}).
And, 
the X-ray spectrum of \WRS definitely has a thermal origin as indicated
by the presence of various spectral lines that originate in thermal plasma 
and this is also the case at high energies.  \citet{sk_07} detected a 
Fe XXV K-line at 6.67 keV in the {\it XMM-Newton} spectra of \WR and 
the zeroth-order HETG data undoubtedly showed that this line is 
associated with \WRS, the WN8 star in the binary system \citep{zhp_10}.

\begin{figure}[hp]
 \centering\includegraphics[width=2.in,height=1.5in]{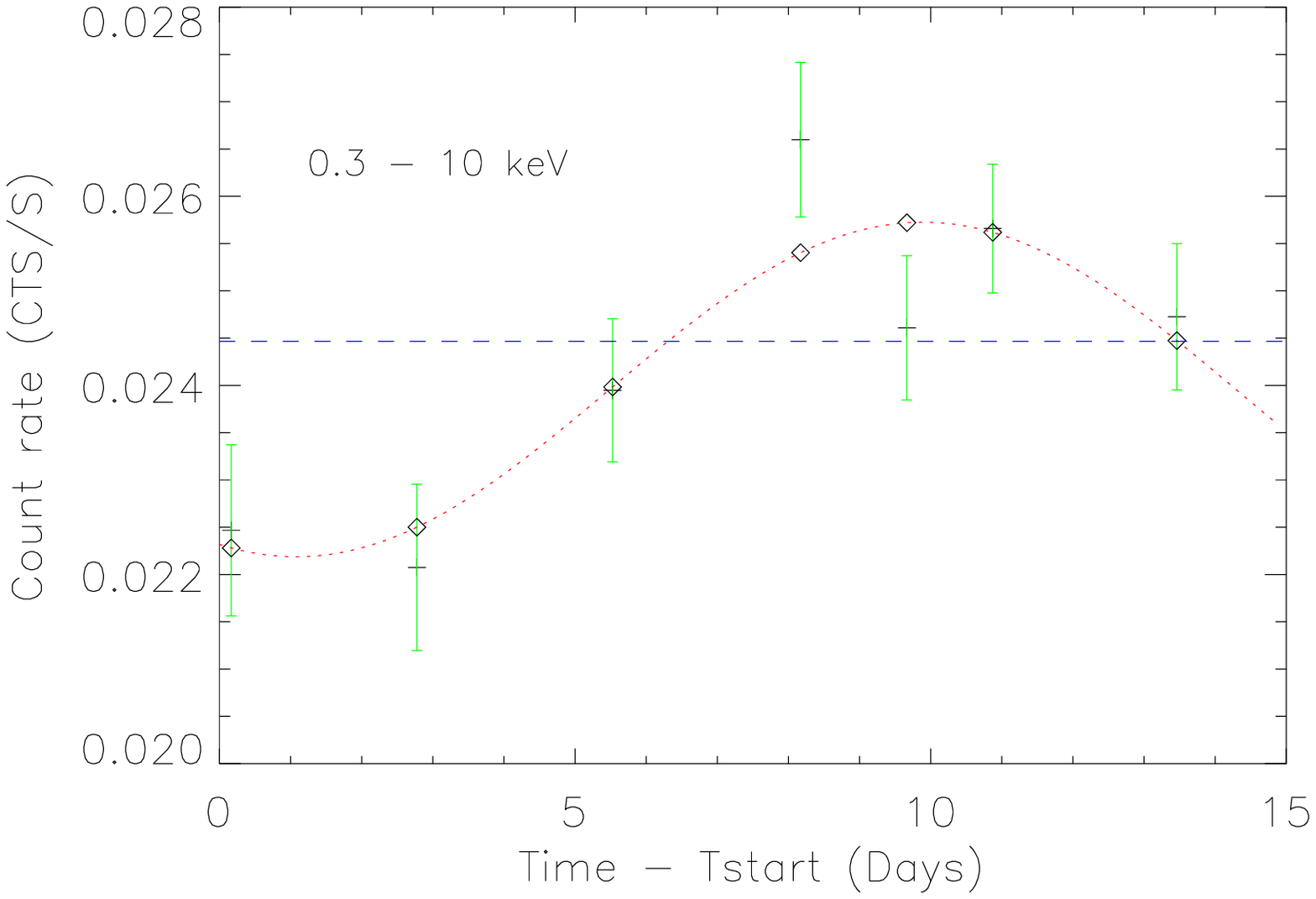}
 \centering\includegraphics[width=2.in,height=1.5in]{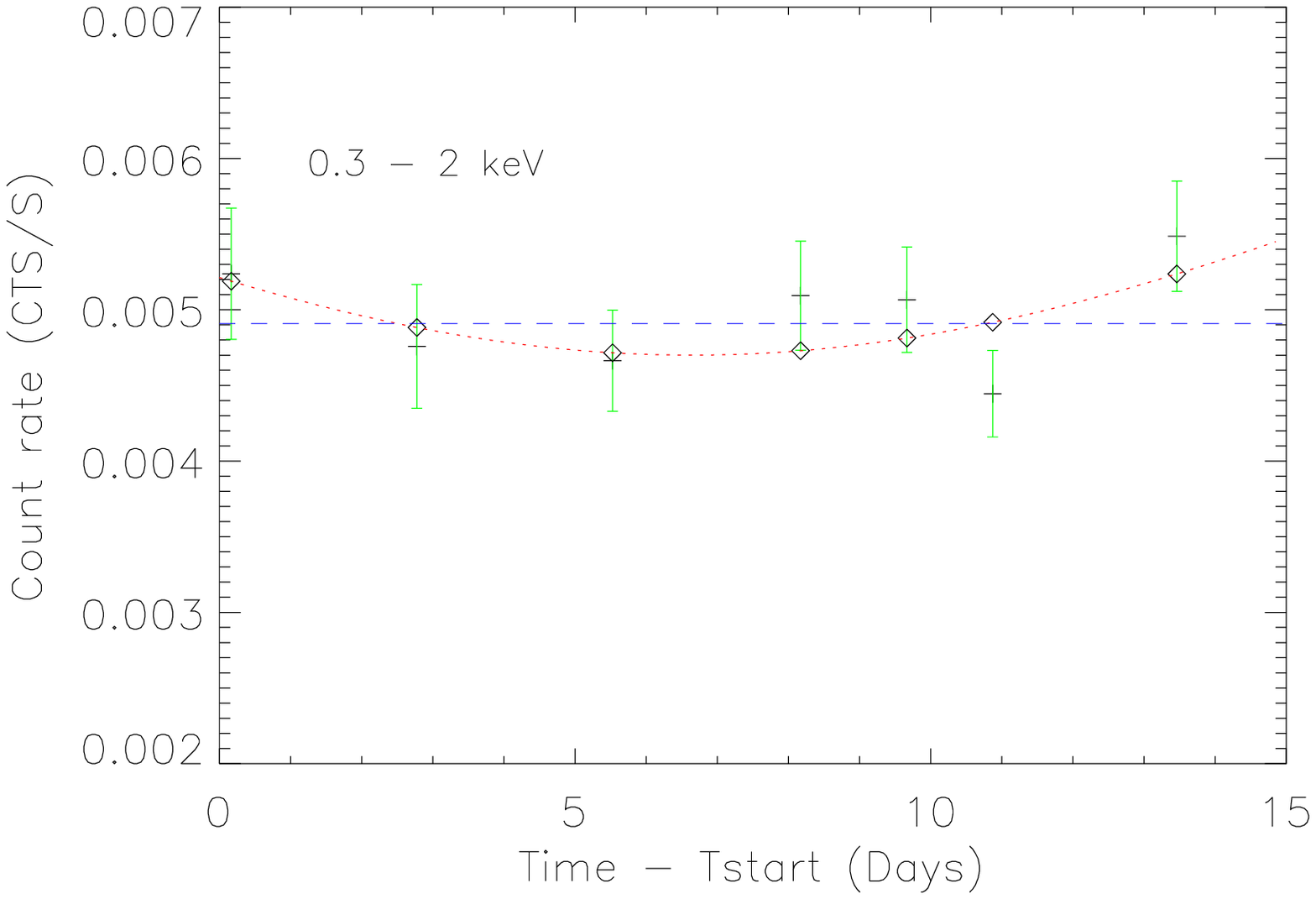}
 \centering\includegraphics[width=2.in,height=1.5in]{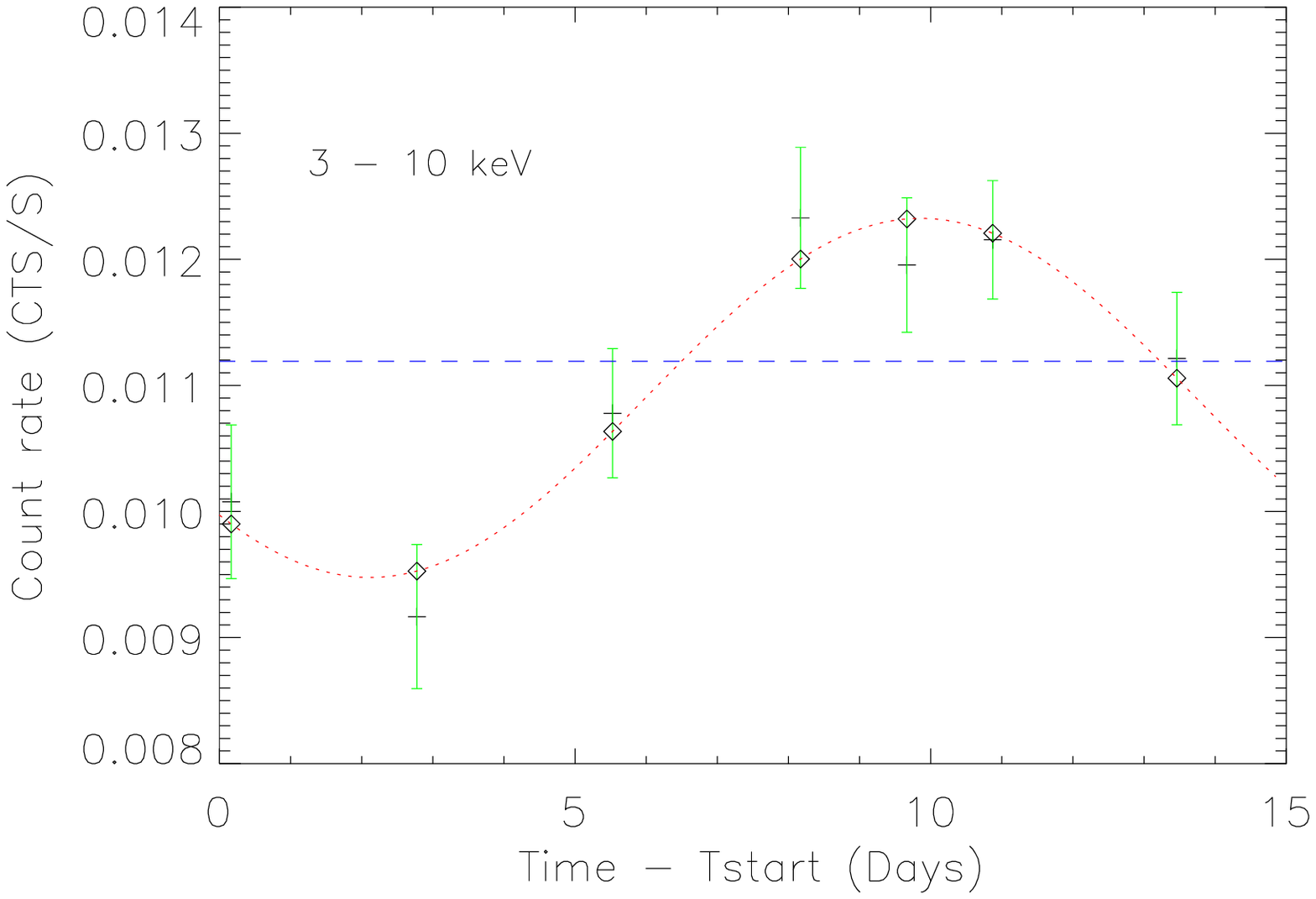}
\caption{
The zeroth-order  background-subtracted light curves of \WR:
the seven data points
correspond to the average count rate for each observation (the data
for ObsIDs 10897 and 10678 were combined, since this is one
observation split into two parts).
The constant-flux fit is presented by the dashed line and the fit
with simple sinusoidal curve is given by the dotted line.
The time (x-axis) is in days and is with respect to the start of the 
first observation in the HETG data set.
}
\label{fig:LC}
\end{figure}

But, the case of X-rays from the stellar wind shocking onto a close
companion seems to find some support from our deep {\it Chandra}
observations that span an almost two-week period. Figure~\ref{fig:LC}
shows the light curve of \WR in different energy bands. Obviously,
a long-term variability is present in the data with a tentative
period of 15-20 days. This variability is well established at high
energies, thus, it should be associated with \WRS which dominates
the X-ray emission at $\mbox{E} > 3$~keV.
We fitted the light curve with two models: a constant flux and a
simple sinusoidal curve.
The values for the reduced $\chi^2$ for various fits are:
$\chi^2(0.3-10~keV) =$ 1.55 (sine) and 3.79 (const);
$\chi^2(0.3-2~keV = $ 1.63 (sine) and 1.14 (const);
$\chi^2(3-10~keV) = $ 0.49 (sine) and 4.49 (const).
The formal goodness of fit is:
0.20 (sine) and 0.0009 (const) for the (0.3-10~keV) LC;
0.18 (sine) and 0.34 (const) for the (0.3-2~keV) LC;
0.69 (sine) and 0.0002 (const) for the (3-10~keV) LC.
The corresponding periods are:
P(0.3-10 keV)$= 17.61\pm3.74$~days;
P(3-10 keV)$= 15.48\pm1.91$~days.
The amplitudes of the variability with respect to the constant flux
are:
7\% in (0.3-10 keV); 13\% in (3-10 keV).

Thus, if the X-rays in \WRS originate from stellar wind shocking onto
a close companion, the observed variability might simply correspond to
the orbital period of the companion star. The changes in the X-ray
emission could be a result from the ellipticity of the orbit (variable
emission measure) and/or they may be due to a variable X-ray 
absorption depending on the inclination angle of the orbit. 
Unfortunately, the quality of the individual spectra in the HETG data 
set (zeroth and first order spectra) does not allow us address these 
issues. 
%
We only note that for an adopted value of 15.48 days for the
orbital period and a total mass of the system of 20~M$_{\odot}$, 
the Kepler's third law 
($a = 0.01952 P^{2/3}_d [M/M_{\odot}]^{1/3}$~au; $P_d$ is the
orbital period in days; $M$ is the total mass) gives $a =
0.33$~au for the semi-major axis of the orbit. For a distance of 630
pc to \WR \citep{ch_92}, the separation between the WN8 star and the
normal star companion will be $\approx 0\farcs0005$. 
Being that close to a luminous hot star  makes it highly unlikely for 
such an object be detected by imaging techniques.
On the other hand, we can speculate that this variable X-ray source
provides extra ionizing photons that are capable of changing the
population in highly ionization stages of elements as carbon, nitrogen 
and oxygen. Given the spectral subtype of the WR star in \WR (WN8),
we can propose that variability in the profiles of NIV and NV lines 
can be expected in the optical with a periodicity as seen 
in X-rays, an effect similar to the Hatchett-McCray effect in massive 
X-ray binaries \citep{hat_77}.

\section{Global Spectral Models}
\label{sec:global}
A systematic approach to modeling the 
total X-ray spectrum of WR 147N$+$S
would be to fit the entire observed spectrum using a reasonable
physical model. We note that the HETG first-order spectra represent
the total (combined) X-ray emission of the two X-ray sources \WRN and
\WRS.
As discussed in \S~\ref{subsec:wr147n}, the most realistic assumption 
for the X-rays in \WRN is that they originate from the CSW region in 
the binary system. Thus, we explore this picture in some detail.
On the other hand, the origin of the X-ray emission in \WRS is not well
constrained and we consider two limiting cases: a distribution 
of adiabatic shocks and a distribution of equilibrium plasma.

\begin{figure}[hp]
 \centering\includegraphics[width=3.20in,height=2.286in]{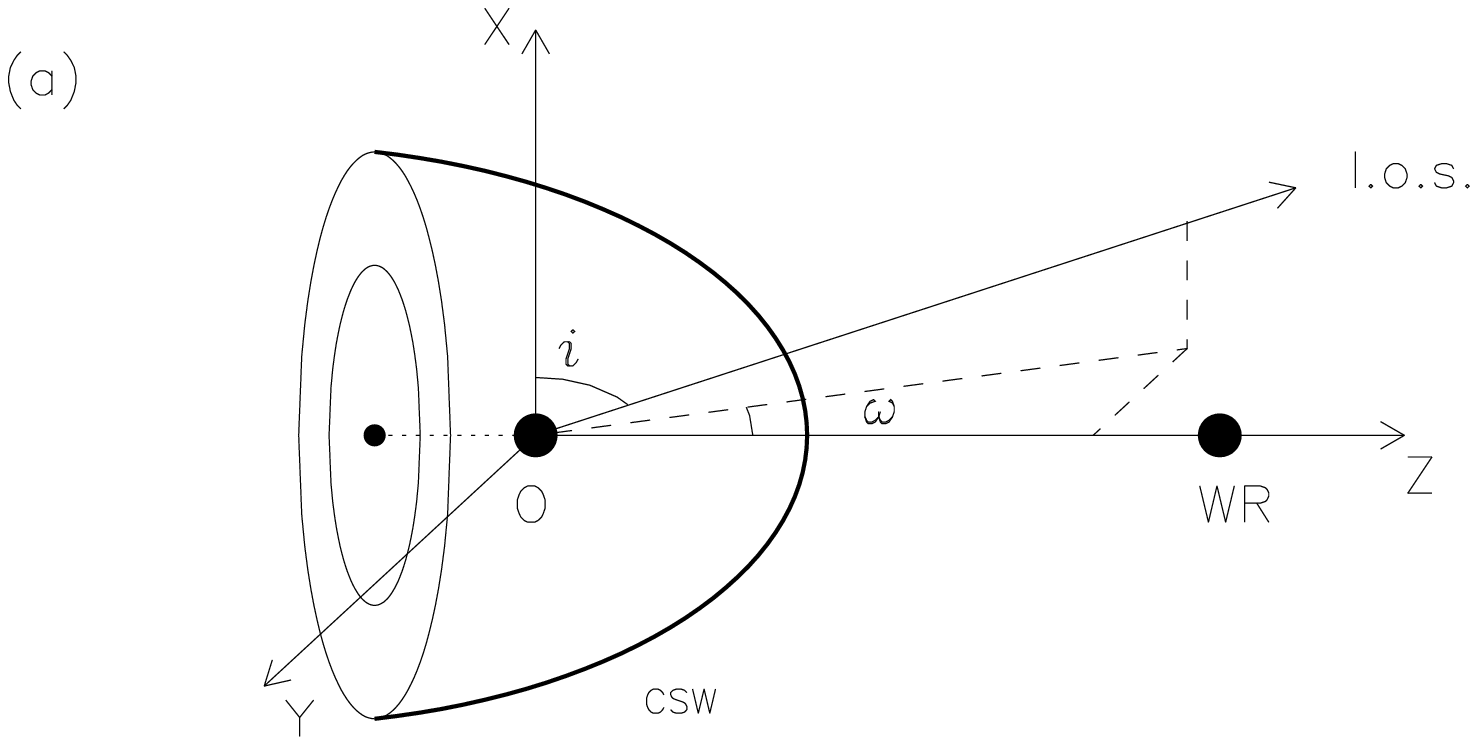}
 \centering\includegraphics[width=3.20in,height=2.286in]{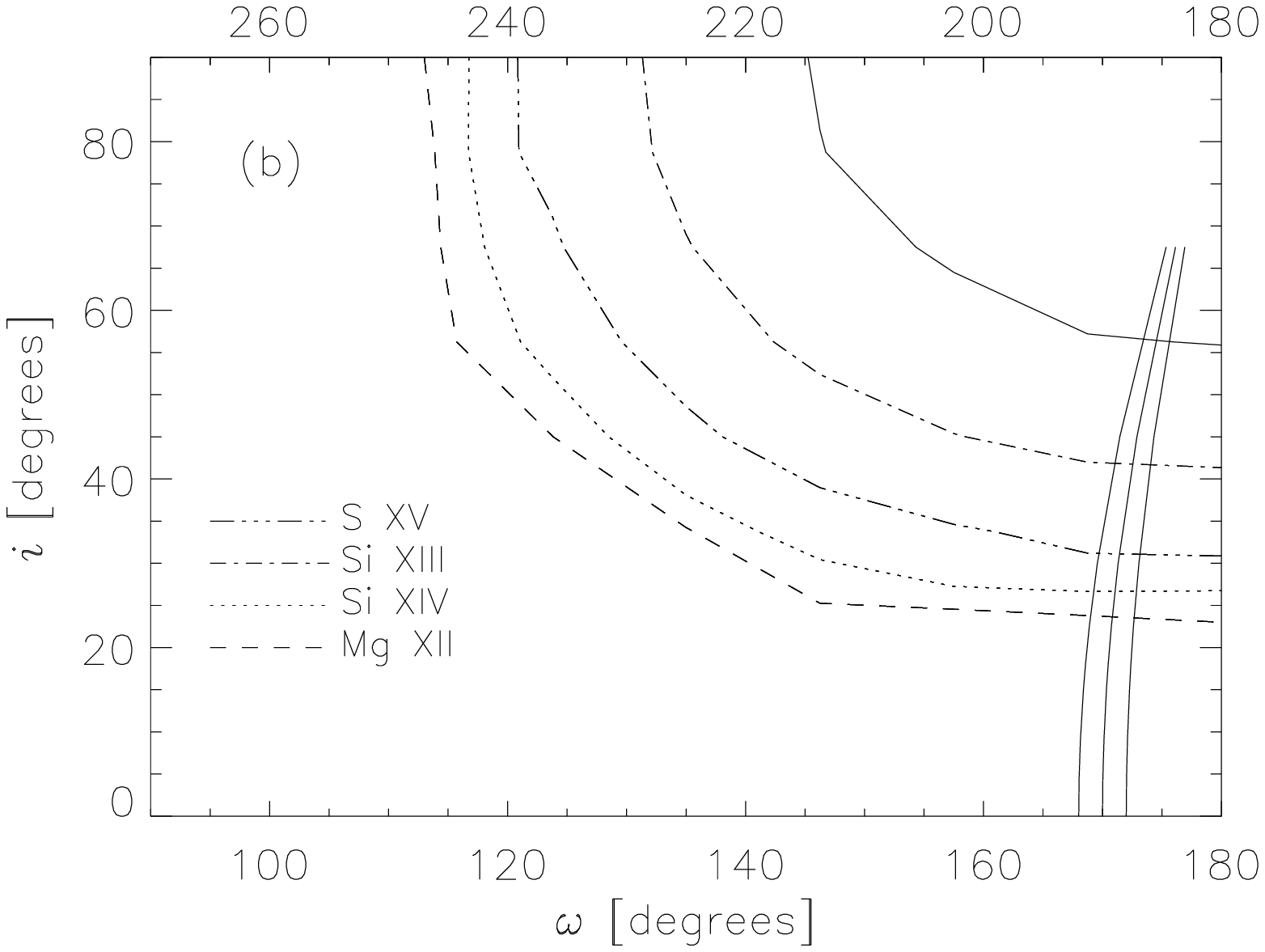}
\caption{
Line shifts of spectral lines that originate in the colliding stellar
wind region. 
{\it Panel (a)} shows
a schematic diagram of the stellar wind interaction in
a WR$+$O binary system. The wind interaction cone 
is denoted by CSW 
(the axis Z is its axis of symmetry; 
the axis X is perpendicular to the orbital plane;
the axis Y completes the right-handed coordinate system);
the line of sight towards observer by l.o.s.; and 
the two related angles, $i$ (orbital inclination) and $\omega$
(azimuthal angle) are marked as well.
{\it Panel (b)} shows the isolines for the observed line shifts for 
various strong spectral lines (Table~\ref{tab:lines}) on the grid of 
theoretical CSW models
with the same stellar wind parameters but for different values of the
inclination and azimuthal angles. The isoline shown with the solid
line gives the upper limit to the line shifts. The three vertical
solid lines correspond to the observed position angle of the WR-O
star axis on the sky (the PA value $\pm1\sigma$ error;
\citealt{nie_98}).
}
\label{fig:grid}
\end{figure}

For the spectrum of \WRN, we adopted the CSW model with non-equilibrium
ionization by \citet{zh_07} and we made a new version for XSPEC that
explicitly takes into account the line broadening (bulk gas
velocities) from the hydrodynamic CSW model. An important feature of 
the new version is that it includes a spatial `offset' for the CSW 
source with respect to a near by source which is the case with the HETG 
first-order spectra of \WRN. 
Moreover, to get realistic line profiles we need to know the position of 
the observer (line of sight) with respect to the CSW region (its axis of 
symmetry). This requires two more model parameters: 
the inclination angle, $i$ (the angle between the line of sight and 
the orbital axis) and the azimuthal angle, $\omega$ (defining the 
position of the observer in the orbital plane). We recall that the CSW 
model by \citet{zh_07} makes use of the hydrodynamic model of 
\citet{mzh_93} which adopts a convention that the O star in the binary 
system is located at the origin of the coordinate system. Panel (a) in
Fig.~\ref{fig:grid} shows a schematic diagram of the wind interaction
of two spherically symmetric stellar winds in a WR$+$OB binary.

Since the CSW spectral calculations require 3D integration, it is not 
feasible to fit for the values of the orbital inclination and
azimuthal angle of the observer's line of sight.
Some information about these parameters comes from the analysis 
of the spectral line profiles (\S~\ref{sec:lines}). We recall that
blue-shifted lines were found in the spectrum of \WRN. This indicates
that the CSW region is inclined towards the observer. 
That is, the CSW region   is located
between the observer and the WN8 star (\WRS).

To explore this in detail, we have run a grid of CSW models for
$\omega \in [90, 180]$. One should keep in mind that due to the axial
symmetry of the CSW region, the line shifts (and line profiles) from
the CSW region are the same for  $\omega \in [90, 180]$ and 
$\omega = 360 - \omega$, and a given value of $i$. 
To be as close as possible to the observational situation, we carried 
out the CSW spectral simulations in XSPEC using the ancillary response 
functions from our HETG observations. We fitted the spectral lines in 
simulated CSW spectra in a similar way as for the real data analysis: 
using a sum of two and three Gaussians for the lines of H-like 
doublets and He-like triplets, respectively.

Panel (b) in Fig.~\ref{fig:grid} shows the isolines for the
observed line shifts of S XV, Si XIV, Si XIII and Mg XII
(from Table~\ref{tab:lines}) in the ($i, \omega$) parameter space 
from our model calculations.
From this in conjunction with the result about the orientation of
the WR-OB binary axis on the sky \citep{nie_98}, we conclude 
that the orbital inclination is smaller than $60^\circ$  with an average
value of $30^\circ$: the isolines 
cluster around $20^\circ-40^\circ$.
The upper limit of the inclination angle comes from the largest
possible value from the $1\sigma$-error on the line shift of 
acceptably well constrained data: in this case it is from the Si XIII 
triplet and is shown with a solid curve in panel (b) of 
Fig.~\ref{fig:grid}. We note that the value for the $1\sigma$-error
of the S XV triplet is beyond this limit but this is likely due to the
poorer quality of the data for this line and the fact that the
contribution from the CSW region is small (Table~\ref{tab:lines}).
This prevents having a tighter constraint (smaller uncertainties) on 
its line shift although the derived value is consistent with those for 
the other lines in our analysis.
From the values of the observed position angle of the WR-OB star axis 
on the sky \citep{nie_98}, we derive a value for the azimuthal angle 
$\omega = 170^\circ-174^\circ$ (see Fig.~\ref{fig:grid}).

\citet{con_99} derived a 
($45^\circ \pm 15^\circ$) range of possible
inclination angles for the \WR system from the analysis of the
morphology of the non-thermal radio source \WRN.
Thus, in the framework of the CSW picture, the results from the HETG 
observations are generally consistent with those from the radio
observations of \WR. But, we now obtain an additional piece of
information: the CSW region is inclined
towards the observer.
Blue-shifted lines were detected, so the CSW region lies between
the observer and the WN8 star.

Our global spectral model consists of two components: a CSW NEI model
representing the X-ray emission from \WRN and a thermal plasma component
for the X-ray spectrum of \WRS. In accord with the spectral analysis
of \citet{zhp_10}, both spectra are subject to the same
X-ray absorption from the interstellar matter (ISM; N$_H =
2.3\times10^{22}$ cm$^{-2}$) and there is an excess
X-ray absorption of the \WRS spectrum which we assume is due to the
WN8 wind. Since the value for the ISM absorption corresponds well to
the optical extinction to \WR, we keep its value fixed 
in this analysis. The WN-wind absorption is a free parameter and for
that purpose we adopted the 
cold wind approximation by making use 
of the {\it vphabs} model in XSPEC. This allows us to impose the same
set of chemical abundances on the hot X-ray emitting plasma and the
wind absorber. We recall that the CSW model at hand takes into account
that the shocked OB-star wind has solar abundances while the shocked WN
wind has a different set of abundances that can be adjusted in the
fitting process. Thus, the CSW model and the model for the X-ray
emission from \WRS share their abundances.
For consistency with the previous studies we adopted
the same set of WN abundances (see \citealt{sk_07}; \citealt{zh_07})
as only the values for Ne, Mg, Si, S, Ar, Ca and Fe were allowed to
vary while those for the other elements were held fixed in the fits.

The CSW model uses nominal stellar wind parameters
(V$_{WR} = 950$~km~s$^{-1}$, $\dot{M}_{WR} = 4\times 10^{-5}$
M$_{\odot}$~yr$^{-1}$;
V$_{O} = 1600$~km~s$^{-1}$, $\dot{M}_{O} = 6.6\times 10^{-7}$
M$_{\odot}$~yr$^{-1}$;
$[\dot{M}_{O} V_{O} / \dot{M}_{WR} V_{WR}] = 0.028$) and a value of 
403 au for the projected binary separation (\S~\ref{sec:thesystem}).
But the mass-loss values for the winds 
of both stars were reduced by a factor of 4
to correct for the CSW luminosity discrepancy revealed from the 
analysis of the undispersed spectra \citep{zhp_10}. The orbital 
inclination and azimuthal angle were held fixed to
$i = 30^\circ$ and  $\omega = 171^\circ$.
Also, a spatial 
offset with a fixed value of 
$\Delta = 0\farcs603$ (\S~\ref{sec:lines})
was adopted for the CSW X-ray source (\WRN).

As mentioned above, we adopted thermal plasma models for the X-ray
spectrum of \WRS that consider two limiting cases. The first case is
a distribution of adiabatic shocks with non-equilibrium ionization
effects (NEI shocks) and we made use of our custom model
for XSPEC which was successfully used in the analysis of the X-ray
spectra of SNR 1987A (e.g. \citealt{zh_09} and references therein).
The second one considers a distribution of emission measure of thermal
plasma in ionization equilibrium (CIE plasma) and we adopted our 
custom model which is similar to $c6pvmkl$ in XSPEC but uses the 
$apec$ collisional plasma model for the X-ray spectrum at given plasma 
temperature. We 
note that the line broadening is self-consistently 
calculated in the CSW model and for the spectrum of \WRS we adopted 
Gaussian broadening using the $gsmooth$ model in XSPEC.

\begin{figure}[hp]
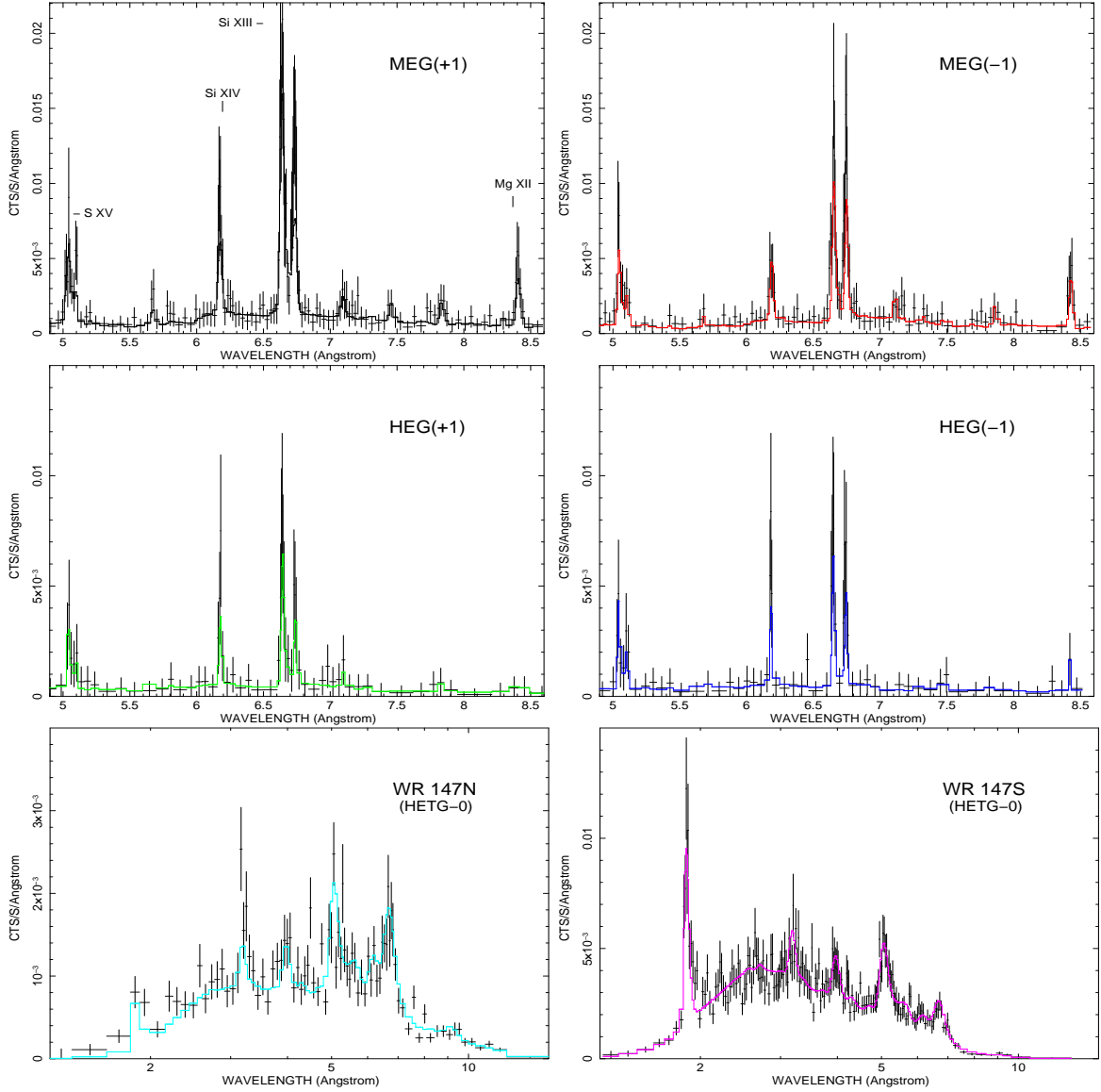

 \centering\includegraphics[width=2.in,height=3.in,angle=-90]{f5a.eps}
 \centering\includegraphics[width=2.in,height=3.in,angle=-90]{f5b.eps}
 \centering\includegraphics[width=2.in,height=3.in,angle=-90]{f5c.eps}
 \centering\includegraphics[width=2.in,height=3.in,angle=-90]{f5d.eps}
 \centering\includegraphics[width=2.in,height=3.in,angle=-90]{f5e.eps}
 \centering\includegraphics[width=2.in,height=3.in,angle=-90]{f5f.eps}
\caption{
HETG background-subtracted spectra of \WR and the two-component model
fit (Table~\ref{tab:fits} and Fig.~\ref{fig:dem}). The first and zeroth 
order spectra were re-binned to have minimum 
10 and 20 counts per bin, respectively.
}
\label{fig:fits}
\end{figure}

\begin{figure}[hp]
 \centering\includegraphics[width=2.80in,height=2.0in]{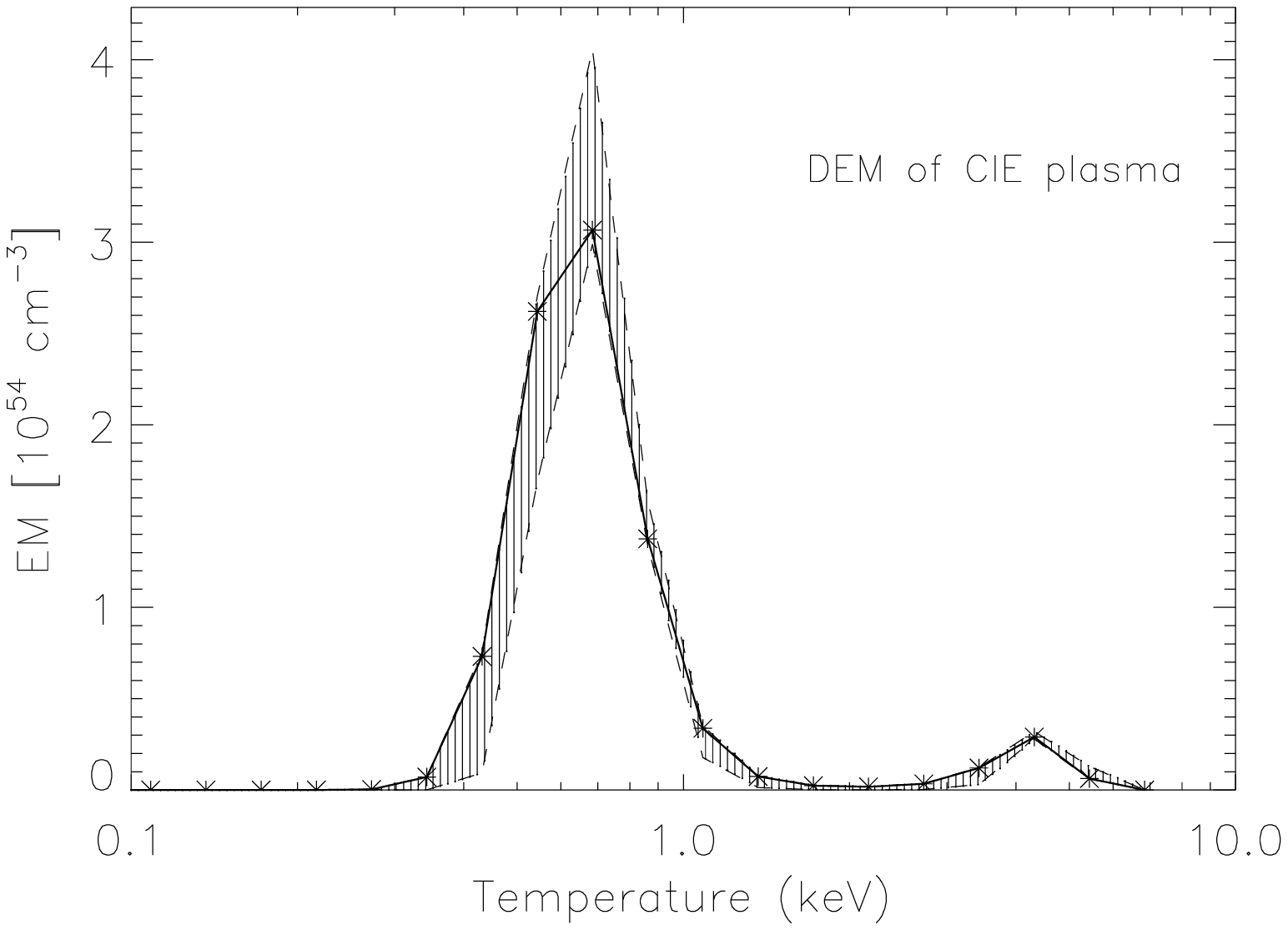}
 \centering\includegraphics[width=2.80in,height=2.0in]{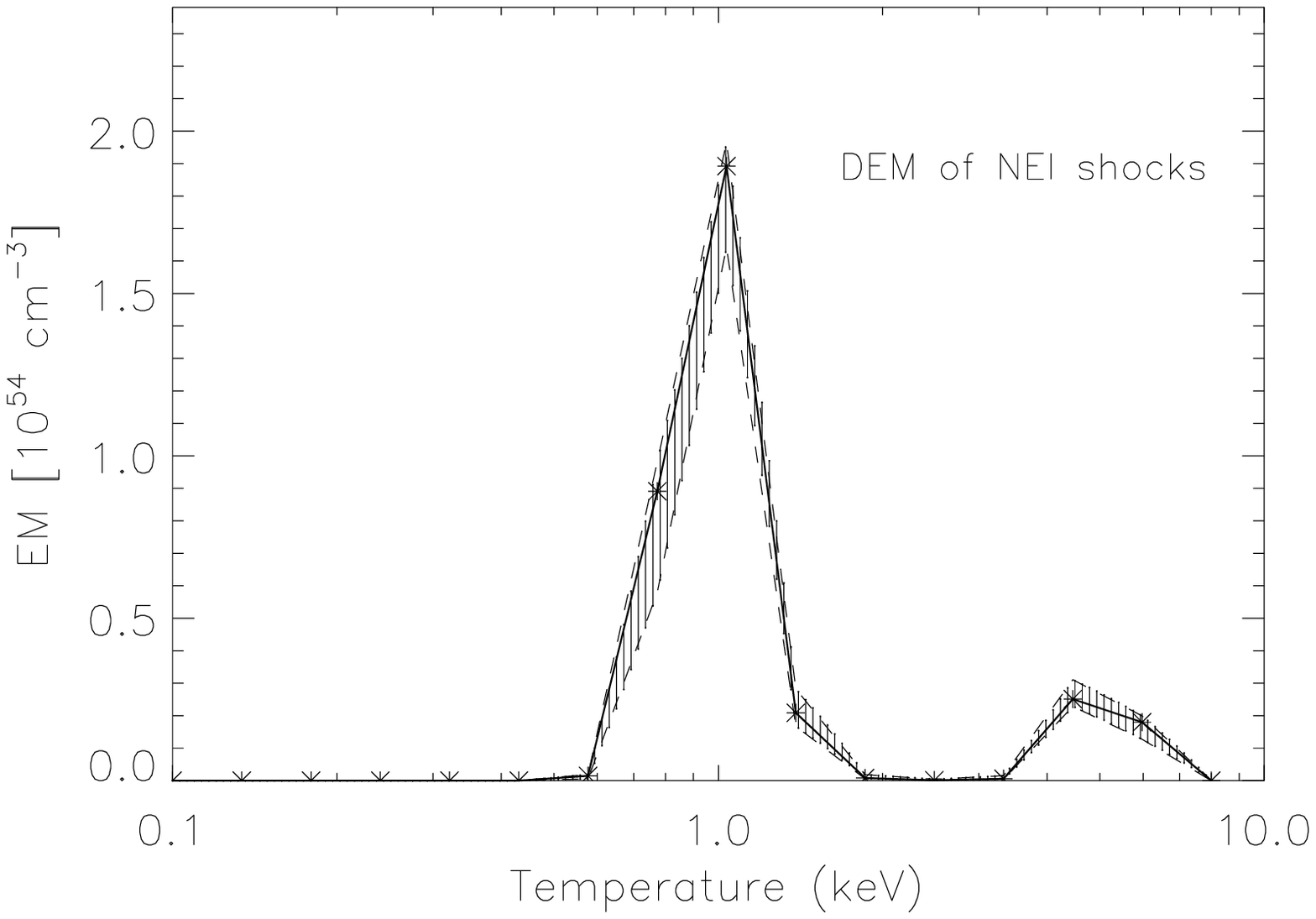}
\caption{
Emission measure (EM) of the \WRS distribution of thermal plasma in
collisional ionization equilibrium (DEM of CIE plasma) and of adiabatic 
shock with non-equilibrium ionization effects taken into account 
(DEM of NEI shocks).
The shaded area represents $1\sigma$ errors from the fits. 
The EM values are for adopted distance of 630 pc to \WR.
}
\label{fig:dem}
\end{figure}

We fitted simultaneously four HETG first-order spectra of \WR:
MEG($+1$), MEG($-1$), HEG($+1$), HEG($-1$); and two zeroth-order 
(undispersed) spectra: one for \WRN and \WRS, respectively.
Table~\ref{tab:fits} and Figures \ref{fig:fits} and \ref{fig:dem} 
show the corresponding fit results.
The following results are worth noting. 
We see that there is a good correspondence between the derived
abundances from the models that assume CIE plasma or NEI shocks are
responsible for the X-ray emission of \WRS. 
The total observed X-ray flux of \WR from the {\it Chandra} HETG
is $1.31\times10^{-12}$ ergs cm$^{-2}$ s$^{-1}$ which is $\sim12$\%
smaller than that from the {\it XMM-Newton} observation in November 
2004 \citep{sk_07}. We believe that this difference is due to 
the calibration uncertainties between the two telescopes but the 
15.48-day X-ray variability of \WRS (see Fig.~\ref{fig:LC}) may also 
contribute some part of it. Note that the flux value from HETG is the 
average X-ray flux over the proposed variability period while the
{\it XMM-Newton} data provide a `snap-shot' X-ray spectrum of \WR.
Despite the fact that we adopted an average line broadening for the 
\WRS spectrum, the derived value (FWHM) from the global fits is in
agreement with the results from the fits to individual lines (see
Table~\ref{tab:lines}).
But perhaps the most interesting and robust result from the global 
fits is that all the adopted models require  very hot plasma to be
present in \WRS (Fig.~\ref{fig:dem}). We mention that the hot plasma 
(CIE or NEI) at temperatures kT$~> 3$~keV supplies $\sim 56-63$\% of 
the total observed flux from \WRS.

Finally, we note that we have run two more cases of the global spectral
models that have orbital inclination and azimuthal angle fixed
correspondingly to: (i) $i = 10^\circ$ and  $\omega = 170^\circ$;
(ii) $i = 50^\circ$ and  $\omega = 174^\circ$. Their results are
consistent with those in
Table~\ref{tab:fits} and Figures \ref{fig:fits} and \ref{fig:dem}
of our basic case with $i = 30^\circ$ and  $\omega = 171^\circ$.






\section{Discussion}  
\label{sec:discussion}
Based on the zeroth-order data \citep{zhp_10} and the analysis of the
first-order spectra presented in this study, we have argued that the
northern X-ray source, \WRN, is likely  associated with the CSW region 
of the \WR binary system, 
while we speculated that the X-ray emission of its southern 
counterpart,  \WRS, is likely due to stellar wind shocking onto a 
close unseen companion. 
Next, we discuss in some detail two of the most important results
related to this physical picture.

\subsection{Reduced Mass-Loss Rates}
As described in \S~\ref{sec:global}, we reduced the mass-loss rates by
a factor of 4 to calibrate the CSW X-ray luminosity to that of
\WRN. On the other hand, the amount of absorption due to the
stellar wind can provide us with another estimate of this parameter.
From the radial column density of the 
assumed `cold' stellar wind and adopted
\WR abundances, we derive the corresponding minimum value of the
absorption column density for the X-ray source located in the wind at 
some distance from the star: 
N$_{H, wind} = 2.76\times10^{22} \dot{M}_5 r_{12}^{-1} V_{1000}$
cm$^{-2}$, where 
$\dot{M}_5$ is the WN8 mass loss in units of $10^{-5}$ \dotM; 
$r_{12}$ is the distance from the star in $10^{12}$ cm;
$V_{1000}$ is the wind velocity in units of 1000\kms.
If the X-ray source in \WRS is located at distance equal to the
semi-major axis of the orbit of the hypothesized close companion 
(\S~\ref{subsec:wr147s}; 
$a = 0.33~\mbox{au} \approx 5\times10^{12}$~cm), we see that the
minimum column density of the stellar wind is $\sim 2-3$ times
larger than the value derived from the global fits
(Table~\ref{tab:fits}) even for the factor of 4 reduced mass-loss rate
($\dot{M}_5 = 1$).
A likely explanation for such a discrepancy is that we adopted a
`cold' stellar wind absorption model in the global spectral fits, that
is all the chemical elements are in their neutral state. 
Cold gas is an efficient X-ray absorber, so, this model underestimates 
the column density of the stellar wind. In reality, the stellar wind 
is ionized considerably, more so in vicinity of the WN8 star where 
the putative close companion is orbiting the primary star.
This results in a higher X-ray transparency of the ionized wind.
Thus, to have the same optical depth as in the case of the
`cold' wind, the ionized wind must have a larger
effective column density.
This can likely resolve the wind-absorption discrepancy.
However, to test such an absorption model 
we need to know the exact ionization structure of the stellar wind. 
This requires a detailed modeling in the optical-UV spectral range 
which is beyond the scope of the present study. We only note that
we find higher values for N$_{H, wind}$ from the fits, with no
deterioration of their quality, if we
enforce a smaller column density for H, He and N, thus mimicking an
appreciable deviation from neutral state for these elements.
On the other hand, the important conclusion is that both the CSW
luminosity and the WN8 stellar wind absorption indicate a reduced mass
loss of the WN8 star by a factor of $\sim 4$ compared to its nominal
value for \WR (\S~\ref{sec:thesystem}). 

From a detailed spectroscopy study of the optical-to-infrared spectrum
of \WR, \citet{mo_00} concluded that 
the clumping factor in the stellar wind 
of the WN8 star is in the range 0.04 to 0.25,  suggesting a
reduction to the mass-loss rate of 0.2 - 0.5 compared to the values
derived under assumptions of wind homogeneity. Thus, if the smaller
mass loss of the WN star needed to explain the X-ray data is due to
wind clumping, this would mean that a considerably inhomogeneous
wind extends out to  large distances from the star (e.g., to the 
thermal radio emission region of $\sim 1000$ stellar radii) and only 
further beyond does it become homogeneous and form the CSW region 
of \WR. 

But an alternative and interesting explanation can be proposed as
well. A homogeneous stellar wind can have a smaller mass-loss rate,
provided the distance to \WR is less than the currently adopted
value. We note that  for the mass-loss values derived from the radio 
observations $\dot{M} \propto d^{1.5}$ (\citealt{pf_75}; 
\citealt{wb_75}).
It is worth noting that the distance to \WR may bear some
appreciable uncertainties since it was derived by \citet{ch_92}
from comparative NIR photometry with the only known galactic
WN8 star (WR 105) which is a member of an association and
WR 105 is now re-classified as a WN9h star \citep{vdh_01}.
Moreover, due to the appreciable uncertainties in the spectral type of
the OB companion, \citet{lepine_01} pointed out another discrepancy
related to the optical emission from \WR: either the WR star is too
bright (by 1.5 mag) or its OB companion is too faint (by
1.5 mag) for their respective spectral types.
We believe that future spatially resolved observations of \WR in the
optical can help us obtain an accurate distance to this object based
on a well-constrained spectral type and luminosity class of the OB
companion in the binary system. In turn, the mass-loss rate of the 
WN8 star will be constrained better which could help us reveal the 
importance of clumpiness in the stellar wind at large distances from 
the star.
Also, a well-constrained luminosity of the OB star will result in a
better estimate of its X-ray luminosity, using the L$_X$ - L$_{bol}$
relation for hot massive stars. Consequently, we will have a more
realistic value for the contribution from the OB star to the total
X-ray emission of \WRN. Note that the higher the L$_X$ of the OB star,
the higher its contribution to the X-ray emission of \WRN, thus,
the higher the mass-loss reduction required.
And all this may allow us build a 
self-consistent physical picture of the CSW binary \WR.

\subsection{High Plasma Temperature in \WRS}
The presence of very hot plasma (kT$~\sim 4$~ keV;
Fig.~\ref{fig:dem}) in the X-ray emission region of the WN8 star (\WRS) 
is a result that does not depend on the accuracy of the distance to the 
observed object. We recall that in the framework of the CSW model the 
shape of the X-ray spectrum of \WRN is well matched using the currently
accepted values for the wind velocities (\S~\ref{sec:thesystem}).
But the high plasma temperature in \WRS poses some problems for the
\WRS emission model: X-rays from stellar wind shocking onto a 
close companion.
Given the chemical composition of the WN8 wind (Table~\ref{tab:fits}),
the temperature of the shocked plasma is 
$\mbox{kT}_{sh} = 2.27 V_{1000}^2$~keV, where $V_{1000}$ is the shock 
velocity in units of 1000\kms. Thus, the maximum temperature we
can get for V$_{wind} = 950$\kms is considerably smaller than that
derived from the global spectral fits. 

We think that one way to resolve this problem could be the following.
In  the adopted scenario of X-rays from a stellar wind shocking onto 
a close (normal star) companion, apart from its own velocity the WN8 
wind gets additionally accelerated by the gravitational field of the 
normal star. Qualitatively, we can assume that near the surface of 
the close companion the effective velocity of the WN wind is:
V$_{eff}^2 = \mbox{V}_{wind}^2 + V_G^2$, where 
V$_{G}^2 = 2 G M_{CO}/ R_{CO}$; $G$ is the gravitational constant;
$M_{CO}$ and $R_{CO}$ are the companion mass and radius, respectively.
In such a case, a value of V$_{G} = 930$\kms is required to match a
plasma postshock temperature of $\sim 4$~keV.

To model the X-ray emission from \WRS, we adopted two limiting cases 
of hot plasma distribution: (i) collisional plasma with equilibrium
ionization; (ii) plasma in adiabatic shocks with non-equilibrium
ionization (\S~\ref{sec:global}). \citet{zh_07} discussed the NEI
effects in colliding stellar wind shocks in some detail and 
introduced a dimensionless parameter $\Gamma_{NEI}$. This parameter is
a measure of whether or not the NEI effects are important: if
$\Gamma_{NEI} \leq 1$ they must be taken into account; and they can be 
neglected if $\Gamma_{NEI} \gg 1$. For the scenario of the
X-rays from stellar wind shocking onto a close companion, adopted here
for \WRS, we have $\Gamma_{NEI} \gg 1$ for the stellar wind parameters
of \WR (even with the reduced mass-loss rates) and the semi-major axis
of the companion orbit of $a = 0.33~\mbox{au}$. 

Similarly, we can introduce a dimensionless parameter that can be
an indicator of whether the X-ray emitting plasma is associated with
adiabatic or radiative shocks. Namely, $\Gamma_{cool} =
t_{flow}/t_{cool}$, where $t_{flow} = r/V_{wind}$ and 
$t_{cool} = \frac{3}{2}\frac{nkT}{\Lambda_{cool}}$, 
($\Lambda_{cool}$ is the cooling
function for collisionally ionized  optically-thin plasma).
Note that if $\Gamma_{cool} \ll 1$ the shocks are adiabatic while
values of $\Gamma_{cool} \ge 1$ indicate radiative shocks.
If for simplicity we assume that the cooling is due only to 
bremsstrahlung emission (which gives a lower limit to the cooling,
thus, an upper limit to $t_{cool}$),
for helium dominated plasma with stellar wind parameters 
$\dot{M} = 10^{-5}$\dotM, V$_{wind} = 950$\kms and orbital radius of
the close companion  $a = 0.33~\mbox{au}$, the dimensionless parameter
becomes: $\Gamma_{cool} = 3.64 T_{keV}^{-0.5}$, where $T_{keV}$ is the
plasma temperature in keV. From this and Fig.~\ref{fig:dem} we see
that $\Gamma_{cool} \ge 1$ for the X-ray plasma in \WRS. 

Thus, in the framework of the adopted X-ray production mechanism of
stellar wind shocking onto a close companion, the values of
$\Gamma_{NEI} \gg 1$ and $\Gamma_{cool} \ge 1$ indicate 
that X-ray emission from a distribution of CIE plasma
is a more appropriate physical model for the X-ray spectra of \WRS. 

In such a case and since the shocks are radiative, we have an upper
limit on the available energy (luminosity) that can be converted into
X-ray emission: no more energy is emitted than the energy flux 
crossing the shock front per unit area. 
Taking into account the additional acceleration of the
wind in the
gravitational field of the close companion (see above), we have:
L$_X = \frac{1}{2}\int \rho \left(V_{wind,\perp} + V_G \right)^3 dS$ 
($\rho$ is the density of the wind in front of the shock; 
$V_{wind,\perp}$ is the wind velocity component perpendicular to the 
shock front; $S$ is the shock surface).
For radiative shocks, the corresponding shock surface 
follows the shape of the stellar surface and  we derive: 
L$_X = \frac{1}{8} \left(\frac{R_{CO}}{a}\right)^2 L_{wind}
\left[1 + 3\left(\frac{V_G}{V_{wind}}\right) + 
3\left(\frac{V_G}{V_{wind}}\right)^2\right]$, where 
$L_{wind} = \frac{1}{2}\dot{M} V_{wind}^2$ is the mechanical
luminosity of the WN8 wind.
Note that this is similar to  eq.(80) in \citet{usov_92} but with a
correction factor for the gravity of the companion star.
And, the simple considerations presented here allow us to further check
the consistency of the adopted physical picture. For a distance of 
630 pc to \WR and the value of unabsorbed X-ray flux derived for \WRS 
(Table~\ref{tab:fits}), we have L$_X = 1.25\times10^{33}$ ergs s$^{-1}$.
Adopting the values of $\dot{M} = 10^{-5}$, $V_{wind} = 950$\kms,
$a = 0.33$ au and $V_G = 930$\kms (for the latter see above),
we derive $R_{CO} = 1.6$ R$_{\odot}$ and in turn we use the value 
of $V_G$ to derive $M_{CO} = 3.6$ M$_{\odot}$.
Thus, we see that the physical characteristics of the putative close 
companion of the WN8 star in \WR are consistent with those of an AB 
main sequence star \citep{allen_73}.
Note that the mass and radius of the companion star will be smaller,
provided \WR is located closer to us than it is currently assumed (it
can be shown that in the picture described above 
$R_{CO}, M_{CO} \propto d^{0.25}$).

It is important to emphasize that the quantitative estimates presented
here serve only as a check on the global consistency of the
physical picture we speculated about in the case of \WRS and they do
not represent a detailed physical model. Nevertheless, two other results
deserve a brief discussion in the framework of the adopted physical 
picture.

If the X-rays from \WRS originate from a region close to the
unseen and much cooler companion (therefore, {\it not a strong source} 
of far UV photons) of the main WN8 star, this may qualitatively explain 
why we do not find suppressed forbidden line in the helium-like triplets 
of Si XIII and S XV. We note that the forbidden and intercombination 
lines in these line complexes have intensities that are within 
$(1-2)\sigma$ of their nominal values (see  Table~\ref{tab:lines} and 
compare with the {\it Chandra} ATOMDB). This fact seems to be in accord 
with the conclusion above that the emission of CEI plasma is a more
appropriate model for the X-ray spectrum of \WRS. But, if much better
grating data for this source become available in the future and they
reveal enhanced emission for the forbidden line in Si XIII and/or S XV, 
then this will pose a problem for the physical picture adopted here for 
the X-ray emission from \WRS. 

Grating data with very high quality are also needed to study possible 
variations of the line centroids in the X-ray spectrum of \WRS. Such
variations are expected with the proposed orbital period of the unseen 
companion if its orbit is not seen pole-on from an observer. For the 
data at hand, we can analyze in detail only the integrated over the 
orbital period X-ray spectrum of \WRS. Thus, we can speculate that the 
derived line centroids being consistent with zero line shifts 
(see \S \ref{sec:lines}) may simply
indicate that the axis of the orbital plane of the unseen companion 
in \WRS (the hypothesized third star in the system) is almost along 
our line of sight.

Finally, it is worth noting that the X-ray emission from \WRS (the WN8 
star in the \WR binary system) shares common characteristics with that 
of presumably single WN stars detected in X-rays \citep{sk_10}: their 
spectra are rather hard and a hot plasma with $kT > 2$ keV must be 
present in the X-ray emission region. This could be a sign that
similar mechanism for X-ray production operates in all these objects.
If we assume that this common mechanism is identified as a
stellar wind shocking onto a close companion, then the simple
considerations presented above would suggest that a 
$L_X \propto L_{wind}$
relation might exist for these WN stars as well. Interestingly,
\citet{sk_10} found such a 
trend which, despite the appreciable 
scatter in the data, is stronger than that for the 
commonly adopted relation between the X-ray and bolometric luminosity 
in massive stars (e.g., compare their Figs. 9 and 10). 
We note that such a scatter might simply indicate 
various companion radii and orbital separation in different WN stars.
However, it should be kept in mind that we need additional pieces of
evidence for the presence of a close companion (supposedly a normal 
star) that would come from future X-ray observations (e.g., revealing
periodic variability) or from observations in other spectral domains 
(e.g, see the end of \S~\ref{subsec:wr147s}). 
The physical picture will 
be on  even more solid ground if objects with 
similar characteristics in the X-rays  are found amongst supposedly 
single objects of the other types of Wolf-Rayet stars (WC and WO).
And we note that an WO object with hard X-ray emission is already 
detected: a presumably single WO star WR 142 
(\citealt{oskin_09}; \citealt{kim_10}).

\section{Conclusions}
\label{sec:conclusions}
In this work, we presented the second part of our analysis of the {\it
Chandra} HETG data of the WR$+$OB binary system \WR which was resolved
into a double X-ray source in the zeroth-order HETG images
\citep{zhp_10}. The basic results and conclusions are as follows:

\begin{enumerate}
\item
Profiles of the strong emission lines in the positive and negative
first-order MEG and HEG spectra  were modeled simultaneously. The
results are consistent with the X-ray emission coming from two sources
spatially separated by $0\farcs603^{+0.10}_{-0.08}$. This value is
in  very good correspondence with the results from the analysis of
the zeroth-order image \citep{zhp_10}.

\item
The line profile analysis showed that the centroids of the lines in the
spectrum of \WRN are {\it blue-shifted} and we find no indication of
suppressed forbidden lines in the He-like triplets either in  the
\WRN or in the \WRS spectra.
The latter means that in both sources: 
(i) the X-ray plasma is not located very close to 
strong UV sources, and (ii) densities in the X-ray plasma
do not significantly exceed the critical density.

\item
Our deep {\it Chandra} observations that span a period of about two
weeks show that the southern source in the binary, \WRS, exhibits  
X-ray variability with a period of $15.48\pm1.91$~days and a 13\%
amplitude with respect to its average flux.

\item
%
The northern X-ray source, \WRN, is probably associated with the CSW 
region in the wide WR$+$OB binary system with orbital inclination
 $i = 30^\circ [0^\circ - 60^\circ]$ .
%
The (variable ) X-rays of its southern counterpart, \WRS (the WN8 star),  
likely result from  the WN8 star  wind shocking onto a close companion
(a hypothesized  third star in the system).

\item
Global spectral models having two components, X-ray emission from 
NEI CSW hydrodynamic models for \WRN and optically-thin plasma spectra
of a distribution of hot plasma for \WRS, provide a good match to the
entire set of HETG (first and zeroth-order) spectra.
%
We note that,
first, a factor of 4 reduction of the stellar mass-loss rate 
relative to currently accepted values is required 
to match the X-ray luminosity of \WRN.
Second,  very hot plasma (kT $ \sim 4$ keV) must be
present in \WRS. Note that the velocity of the WN-star wind 
is not high enough to provide such a postshock temperature. Thus, we
propose that the higher velocity value is due to the additional
wind acceleration in the gravitational field of the close companion.

\end{enumerate}




\acknowledgments
This work was supported by NASA through Chandra grant GO9-0013A to
the University of Colorado at Boulder, and through grant G09-0013B to
the Pennsylvania State University.
The authors thank an anonymous referee for 
her/his comments and suggestions.



{\it Facilities:} \facility{CXO (HETG, ACIS)}.

\clearpage



%

\clearpage

\clearpage

\clearpage

\clearpage

\clearpage

\clearpage

\begin{deluxetable}{lcrrrrcc}
\tablecaption{Line Parameters 
\label{tab:lines}}
\tablewidth{0pt}
\tablehead{
\multicolumn{2}{c}{} & \multicolumn{2}{c}{\WRS} &
\multicolumn{4}{c}{\WRN} \\
\colhead{Line} & \colhead{$\lambda_{lab}^{a}$} & \colhead{FWHM$^{b}$} & 
\colhead{Flux$^{c}$} & \colhead{FWHM$^{b}$} & \colhead{Flux$^{c}$} & 
\colhead{Line Shift$^{d}$} & \colhead{$\Delta^{e}$}  \\
\colhead{} & \colhead{(\AA)} & \colhead{(\kms)} &
\colhead{} & \colhead{(\kms)} & \colhead{} &
\colhead{(\kms)} & \colhead{(arcsec)}
}
\startdata
S XV K$_{\alpha}$ & 5.0387 & 
1300$^{+395}_{-265}$ & 15.36$^{+1.92}_{-1.92}$ & 
$< 500$ & 3.84$^{+0.48}_{-0.48}$
              & -152$^{+16}_{-283}$ & $0\farcs54^{+0.13}_{-0.14}$\\
\,\,\,(i/r)$^{f}$ &  &  & 0.32$^{+0.17}_{-0.14}$  &  & 0.32  &  & \\
\,\,\,(f/r)$^{f}$  &  &  & 0.67$^{+0.20}_{-0.14}$  &  & 0.67  &  & \\
Si XIV L$_{\alpha}$ & 6.1804 & 
665$^{+450}_{-300}$  &  2.94$^{+0.51}_{-0.78}$ & 
$< 1050$ &  1.62$^{+0.65}_{-0.44}$
              & -140$^{+180}_{-80}$  & $0\farcs66^{+0.22}_{-0.25}$ \\
Si XIII K$_{\alpha}$ & 6.6479 & 
1200$^{+225}_{-185}$ & 11.43$^{+2.03}_{-2.66}$ &  
670$^{+261}_{-254}$ &  7.84$^{+2.88}_{-1.80}$ 
              & -246$^{+68}_{-68}$ & $0\farcs54^{+0.10}_{-0.13}$ \\
\,\,\,(i/r)  &  &  & 0.04$^{+0.08}_{-0.04}$  &  
               & 0.22$^{+0.17}_{-0.16}$  &  & \\
\,\,\,(f/r)  &  &  & 0.75$^{+0.20}_{-0.17}$  &  
               & 0.66$^{+0.28}_{-0.19}$ &  & \\

Mg XII L$_{\alpha}$ & 8.4192 & 
1550$^{+680}_{-600}$  &  2.04$^{+0.84}_{-0.79}$ & 
$< 1500$ &  1.15$^{+0.50}_{-0.49}$ 
              & -150$^{+167}_{-107}$ & $0\farcs67^{+0.27}_{-0.09}$ 

\enddata
\tablecomments{
Results from simultaneous fits to the line profiles in the MEG and
HEG $(+1/-1)$ spectra with the associated $1\sigma$ errors.
For the He-like triplets,
the ratios of the intercombination to the resonance line (i/r) and of the
forbidden to the resonance line (f/r) are given as well.
}
\tablenotetext{a}{
The laboratory wavelength of the main component.
}
\tablenotetext{b}{
The line width (full width at the half maximum). 
}
\tablenotetext{c}{
The observed total line/multiplet flux in units of $10^{-6}$ 
photons cm$^{-2}$ s$^{-1}$.
}
\tablenotetext{d}{
The line shift of the \WRN component. 
Note that a negative velocity corresponds to a blueshift.
}
\tablenotetext{e}{
The spatial offset between the two sources (\WRN and \WRS). 
}
\tablenotetext{f}{
Due to the quality of the data, it was assumed that the (i/r) and
(f/r) ratios were the same in the spectra of \WRN and \WRS.
}

\end{deluxetable}

\clearpage

\begin{deluxetable}{lcc}
\tablecaption{Global Spectral Model Results 
\label{tab:fits}}
\tablewidth{0pt}
\tablehead{
\colhead{} & \colhead{CIE Plasma}  & \colhead{NEI Shocks} 
}
\startdata
$\chi^2$/dof  & 561/993 & 526/992  \\
N$_{H, ISM}$$^a$ &  2.3 & 2.3  \\
N$_{H, wind}$$^a$ & 0.23$^{+0.01}_{-0.01}$ & 0.19$^{+0.01}_{-0.01}$ \\
Ne   &  0.0$^{+2.7}_{-0.0}$  & 1.3$^{+1.6}_{-1.3}$ \\
Mg   &  4.1$^{+0.9}_{-0.8}$  & 3.3$^{+0.5}_{-0.5}$  \\
Si   &  3.8$^{+0.7}_{-0.3}$  & 3.5$^{+0.2}_{-0.2}$  \\
S~   &  6.1$^{+0.7}_{-0.7}$  & 5.5$^{+0.4}_{-0.4}$  \\
Ar   &  11.6$^{+2.8}_{-2.8}$  & 8.7$^{+1.3}_{-1.3}$   \\
Ca   &  19.4$^{+5.4}_{-5.3}$  & 12.6$^{+2.4}_{-2.4}$  \\
Fe   &  6.6$^{+0.8}_{-0.9}$   & 7.7$^{+0.6}_{-0.6}$  \\
FWHM$^b$   &  840$^{+400}_{-220}$ & 1000$^{+100}_{-260}$\\
$\tau^c$   &   &   1.03$^{+0.17}_{-0.12}$ \\
F$_{X,~WR~147N}$$^c$  & 0.327 ( 9.5) & 0.326 (10.1) \\
F$_{X,~WR~147S}$$^c$  & 0.975 (26.3) &  0.992 (28.1) \\
\enddata
\tablecomments{
Results from  simultaneous fits to the combined
(WR~147N$+$S) first-order spectra and the individual undispersed
(zeroth-order) spectra for \WRN and \WRS.
Errors are the $1\sigma$ values from the fits.
All abundances
are with respect to their solar values (\citealt{an_89}).
The fixed in the fit abundances are:
H$ = 1$, He$ = 25.6$, C$ = 0.9$, N$ = 140$, O$ = 0.9$, and Ni$ = 1$
(for details see \citealt{sk_07}; \citealt{zh_07}).
The corresponding distributions of emission measure vs. electron
temperature of the hot plasma are shown in Fig.~\ref{fig:dem}.
}
\tablenotetext{a}{
The X-ray absorption in units of 10$^{22}$ cm$^{-2}$.
Only the spectrum of \WRS is subject to excess X-ray absorption
that is due to a cold stellar wind (N$_{H, wind}$) with the same 
abundances as the X-ray emitting plasma.
}

\tablenotetext{b}{
The average full width of half maximum for the lines in the 
spectrum of \WRS. Units are \kms.
}

\tablenotetext{c}{
The ionization age ($n_e t$) of the shocks in units of 
$10^{11}$ cm$^{-3}$ s.
}

\tablenotetext{d}{
The observed X-ray flux (0.5 - 10 keV) followed in parentheses
by the unabsorbed value. Units are $10^{-12}$ ergs cm$^{-2}$ s$^{-1}$.
}
\end{deluxetable}




\end{document}